\documentclass{elsart}

\usepackage{graphics} 
\usepackage{graphicx}
\usepackage{epsfig}
\usepackage{amssymb}
\usepackage{lineno}
\usepackage{subfigure}

\begin{document}

\begin{frontmatter}

\title{Majority-vote on directed Erd{\H o}s--R\'enyi random graphs}
 
\author{F.W.S. Lima} 
\address{Departamento de F\'{\i}sica, Universidade Federal do
  Piau\'{\i}\\Teresina - PI, 64049-550, Brazil}
\ead{wel@ufpi.br}

\author{A.O. Sousa}
\address{Departamento de F\'{\i}sica Te\'orica e Experimental, Universidade
Federal do Rio Grande do Norte\\Natal-RN, 59072-970, Brazil}
\ead{aosousa@dfte.ufrn.br}
 
\author{M.A. Sumuor}
\address{Physics Department, Al-Aqsa University \\ Gaza, Gaza Strip, 
P. O. Box 4051, Pallestian Authority}
\ead{msumoor@alaqsa.edu.ps}

\begin{abstract}
Through Monte Carlo Simulation, the well-known majority-vote model has 
been studied with noise on {\it directed} random graphs. In order 
to characterize completely the observed order-disorder phase 
transition, the critical noise parameter $q_{c}$, as well as the 
critical exponents $\beta/\nu$, $\gamma/\nu$ and $1/\nu$ have been 
calculated as a function of the connectivity $z$ of the random graph.
\end{abstract}

\end{frontmatter}

%%%%%%%%%%%%%%%%%%%%%%%
%64.60.Cn 	Order-disorder transformations; statistical mechanics
%of model systems

%05.10.Ln 	Monte Carlo methods 

%64.60.Fr 	Equilibrium properties near critical points, critical
%exponents

%75.10.Hk 	Classical spin models
%%%%%%%%%%%%%%%%%%%%%%
\section {Introduction}
The majority-vote model (MVM) \cite{gray,mario}, a nonequilibrium model 
defined by stochastic dynamics with local rules and with up-down 
symmetry,  defined on regular lattices shows a second-order phase 
transition with critical exponents $\beta$, $\gamma$, $\nu$ -- which 
characterize the system in the vicinity of the phase transition -- 
identical \cite{mario,santos,crochik} with those of equilibrium Ising 
model \cite{ising,critical}. More general, it has been argued that 
the existence of up-down symmetry in two-state dynamic systems implies 
the same critical behavior (same universality class) of the equilibrium 
Ising model \cite{grin,santos,none}.

On the other hand, MVM on the complex networks exhibit different
behavior, i.e., it belongs to  different universality class 
\cite{campos,lima0,lima1,pereira}. Campos {\it et al.} 
investigated MVM on a small-world network \cite{campos}, which 
was constructed using the square lattice (SL) by the 
rewiring procedure \cite{watts}. Campos {\it et al.} found that the critical 
exponents $\gamma/\nu$ and $\beta/\nu$ are different from these of 
the Ising model \cite{critical} and depend on the rewiring probability.
Pereira {\it et al.} \cite{pereira} studied MVM on 
Erd{\H o}s--R\'enyi's (ER) classical random graphs \cite{er,boll}, 
and Lima {\it et al.}  \cite{lima0} also studied this model on 
random Voronoy--Delaunay \cite{VD} lattice with periodic boundary 
conditions. Very recently Lima \cite{lima1} studied the MVM on 
directed Albert--Barab\'asi (AB) network \cite{ba} and contrary to 
the Ising model \cite{ising} on these networks \cite{alex}, the 
order/disorder phase transition is observed in this system. The
calculated $\beta/\nu$ and $\gamma/\nu$ exponents are different 
from those for the Ising model \cite{critical} and depend on the 
value of connectivity $z$ of AB network. The latter was observed
also for {\it undirected} ER random graph \cite{pereira}.

In this paper we study the Majority-vote model with noise 
on {\it directed} Erd{\H o}s--R\'enyi (ER) random graphs  \cite{er,boll}. 
Through Monte Carlo (MC) simulations and standard finite-size scaling 
techniques we determine the critical exponents  for several 
values of the connectivity $z$ of the graph, in order to characterize
the observed order-disorder phase transition.

\section{The Model}

The majority-vote model, on {\it directed} Erd{\H o}s--R\'enyi (ER) random
graphs, is defined \cite{mario,lima0,pereira,er,boll,jff} by a set of "voters"
or spins variables ${\sigma_i=\pm1}$, located on every site (or node)
of a {\it directed} ER graph. The system dynamics is as follows: For
each spin we determine the sign of the majority of its neighboring
spins. With probability $q$, known as the noise parameter, the spin
takes the opposite sign of the majority of its neighbors, otherwise it
takes the same sign. It is important to emphasize that the noise
parameter $q$ plays the role of the temperature in equilibrium systems
and measures the probability of aligning anti-parallel to the majority
of neighbors. The probability $w_{i}$ of a single spin flip is
given by
\begin{equation}
w({\sigma_i})=\frac{1}{2}\biggl[ 1-(1-2q)\sigma_{i}S\biggl(\sum_{\delta
=1}^{k_{i}}\sigma_{i+\delta}\biggl)\biggl]\mbox{,\,with\,}
S(x)= \left\{
\begin{array}{rcl}
\frac{x}{|x|},& \mbox{se} & x \neq 0\\
 0, & \mbox{se} & x = 0
\end{array}
\right.
\label{eq1}
\end{equation}
where the summation is over all $k_i$ sites connected to spin $\sigma_{i}$. 
It is worth to mention that this probability exhibits up-down symmetry, i.e., 
$w({\sigma_i})=w(-{\sigma_i})$ under the change of the spins in the 
neighborhood of ${\sigma_i}$.

It is argued \cite{gray} that on a square lattice there exists two phases 
for sufficiently small $q$, with a formation of an island of up spins on a sea 
of down spins. The size of this island follows a birth--and--death process 
in which the death rate is larger than the birth rate, thus preventing the 
growth of the island and keeping the down spin phase stable. By symmetry, 
there must exist another phase with spins up. However, if the up--down 
symmetry of the spin flip probability is broken, no phase transition is 
expected, as in the Ising model \cite{gray}. In two dimensions this model 
has a ferromagnetic stationary phase for $0<q<q_c$ undergoing a second-order 
phase transition to a paramagnetic phase at $q_c$ ($q_c=0.075$ for a 
square lattice). The static critical behavior is Onsager-like 
\cite{mario,santos,none}, and according to the argument of Grinstein {\it et al}. 
\cite{grin}, its dynamic critical behavior is the same as model Ising.

Indeed, the majority-vote is a particular case of a general class of polling 
models \cite{social1,social2} composed by interacting two-state opinion 
agents. The main result of the most opinion formation models is that the 
dynamical rule leads to an opinion polarization of the whole population 
along one of the two competing opinions \cite{social1,social2,galam1}. 
Recently, the contrarian concept - agents which have the opinion opposite 
to that of the majority - was introduced to take into account some 
peculiar behavior of the agents \cite{galam2}. With a small fraction 
of contrarians, the system no longer leads to total polarization: the 
minority opinion does not disappear from the population. Above a critical 
fraction of contrarians, polarization does not occur, and the two possible 
opinions are shared each by half the agents \cite{galam2,lama}. 

Depending on the nature of the interactions, networks can be directed
or undirected \cite{graph}. In directed networks, the interaction between any two
nodes has a well-defined direction, which represents, for example,
the direction of material flow from a substrate to a product in a
metabolic reaction, or the direction of information flow from a
transcription factor to the gene that it regulates. In undirected
networks, the links do not have an assigned direction. For example, in
protein interaction networks a link represents a mutual
binding relationship: if protein A binds to protein B, then protein B
also binds to protein A. In other words, in undirected networks, if a
node $A$ is linked to $B$, then $B$ must also be linked to $A$, while
in directed ones, if a node $A$ is linked to $B$, $B$ may not be
linked to $A$, but to some node else instead. In the present work, we 
only consider directed graphs.

Starting with $N$ disconnected nodes, the Erd{\H o}s--R\'enyi network are
generated by connecting couples of randomly selected nodes with a
probability $0 < p < 1$ for every pair, prohibiting multiple
connections (i.e., couples of nodes connected by more than one link), 
until the number of edges equals $z$ (connectivity or number of links of a node). 
In the limit $N \rightarrow \infty$, it is found that the tail (high $z$ region) of
the degree distribution $P(z)$ decreases exponentially, which
indicates that nodes that significantly deviate from the average are
extremely rare. The clustering coefficient - which is the probability that 
two neighbors of the same node are also mutual neighbors \cite{watts,cluster} - is 
independent of a node's degree, so $C(z)$ appears as a horizontal line if 
plotted as a function of $z$. The mean path length is proportional to the 
logarithm of the network size, $l \approx log N$ \cite{boll}, which indicates 
that it is characterized by the small-world property.

To study the critical behavior of the model we consider the
magnetization $M$, the susceptibility $\chi$ and the Binder's
fourth-order cumulant, which are defined by

\begin{equation}
M_N(q)=\biggl[\biggl\langle\biggl|\frac{1}{N}\sum_{i=1}^{N}\sigma_{i}\biggl|\biggl\rangle_t\biggl]_s,
\label{eq2}
\end{equation}

\begin{equation}
\chi_N(q)=N[\langle\langle m^2 \rangle_t \rangle_s-\langle\langle m
\rangle_t \rangle_s^2],
\label{eq3}
\end{equation}

\begin{equation}
U_N(q)=1-\frac{\langle\langle m^{4}  \rangle_t
  \rangle_s}{3\langle\langle m^{2} \rangle_t \rangle_s},
\label{eq4}
\end{equation}

\noindent
where $N$ is the number of vertices of the random graph with fixed
$z$, $\langle\ldots\rangle_t$ denotes time averages taken in the stationary
regime, and $\langle\ldots\rangle_s$ stands for sample averages ($20$ samples).

These quantities are functions of the noise parameter $q$ and, in the
critical region, obey the following finite-size scaling relations
\begin{equation}
M=N^{-\beta/\nu}f_{m}(x)[1+ ...],
\label{eq5}
\end{equation}
\begin{equation}
\chi=N^{\gamma/\nu}f_{\chi}(x)[1+...],
\label{eq6}
\end{equation}
\begin{equation}
\frac{dU}{dq}=N^{1/\nu}f_{U}(x)[1+...],
\label{eq7}
\end{equation}
\noindent 
the brackets $[1+...]$ indicate corrections-to-scaling terms, $\nu$,
$\beta$ and $\gamma$ are the usual critical exponents, $f_{i}(x)$ are
the finite size scaling functions with
\begin{equation}
x=(q-q_{c})N^{1/\nu}
\label{eq8}
\end{equation}
being the scaling variable. Therefore, from the size dependence of
$M_N$ and $\chi$ we obtained the exponents $\beta/\nu$ and
$\gamma/\nu$, respectively. The maximum value of susceptibility $\chi$
also scales as $N^{\gamma/\nu}$. Moreover, the value of $q$ for which
$\chi$ has a maximum, $ q_{c}^{\chi_{\rm max}}=q_{c}(N)$,
is expected to scale with the system size as
\begin{equation}
q_{c}(N)=q_{c}+bN^{-1/\nu},
\label{eq9}
\end{equation}
were the constant $b$ is close to unity. Therefore, the equations 
\ref{eq7} and \ref{eq9} are used to determine the exponent $1/\nu$, as well as
to check the values of $q_c$ obtained from the analysis of the
Binder's cumulant (Eq. \ref{eq4}). Finally, we have also examined if the
calculated exponents satisfy the hyperscaling hypothesis
\begin{equation}
2\beta/\nu+\gamma/\nu=D_{\rm eff}
\label{eq10}
\end{equation}
in order to get the effective dimensionality, $D_{\rm eff}$, for
various values of connectivity $z$.

We have performed Monte Carlo simulation on {\it directed} ER random
graphs with various values of connectivity $z$. For a given $z$, we
used systems of size $N=250$, $500$, $1000$, $2000$, $4000$, $8000$
and $16000$. We waited $10000$ Monte Carlo steps (MCS) to make the
system reach the steady state, and the time averages were estimated
from the next $ 10000$ MCS. In our simulations, one MCS is accomplished
after all the $N$ spins are updated. For all sets of parameters, we
have generated $20$ distinct networks, and have simulated $20$
independent runs for each distinct network.

\section{Results and Discussion}

\begin{figure}[!htb]
  \centering
  \subfigure[]{\label{fig1a}%
    \includegraphics[angle=-90,width=0.48\textwidth]{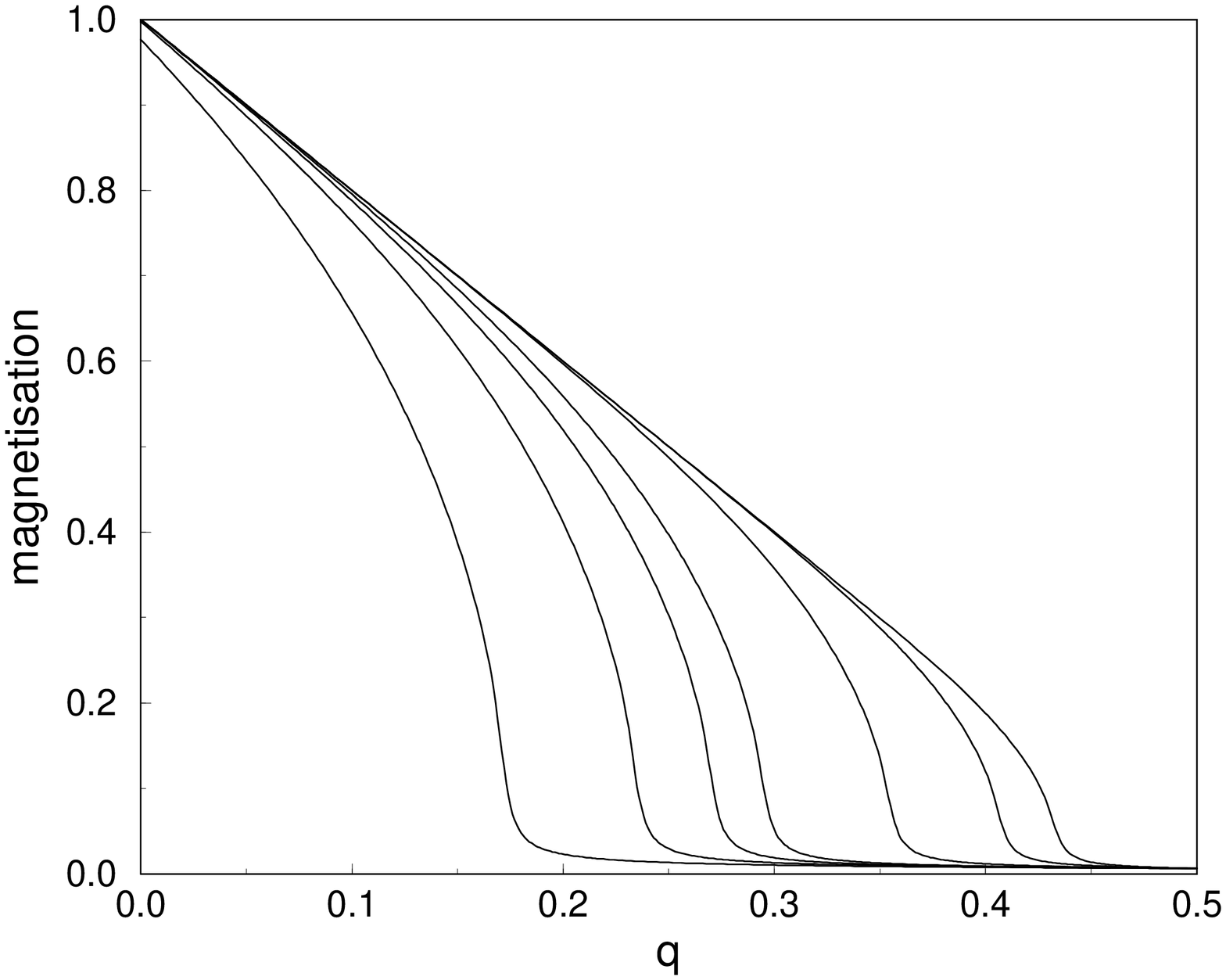}}%
  \quad%
  \subfigure[]{\label{fig1b}%
    \includegraphics[angle=-90,width=0.48\textwidth]{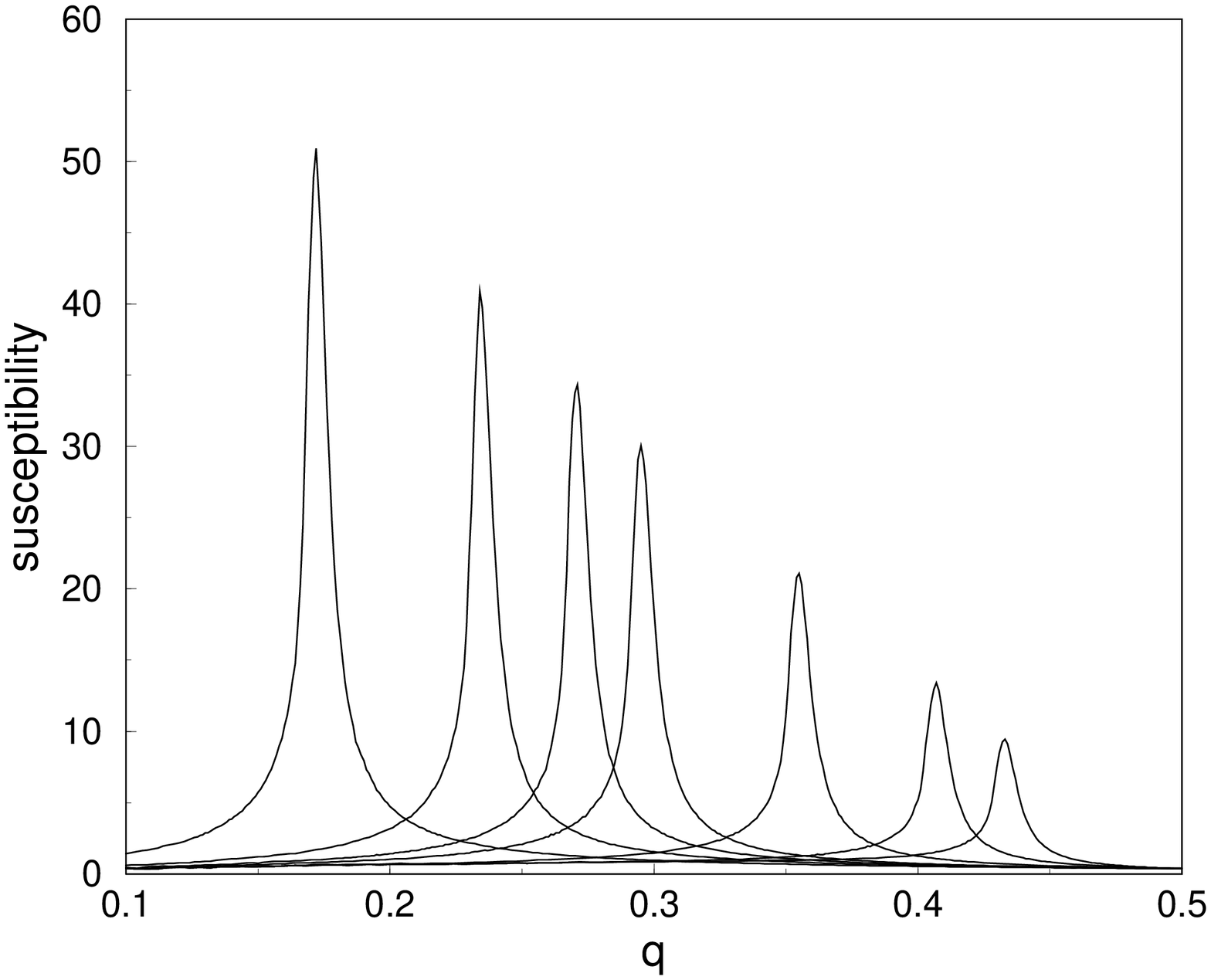}}
\caption{ {\bf (a)} Magnetization and {\bf (b)} susceptibility as a function of 
the noise parameter $q$, for $N=16000$ sites. From left to right we have $z=4$, 
$6$, $8$, $10$, $20$, $50$, and $100$.}
\label{fig1}
\end{figure}

In Fig. \ref{fig1} we show the dependence of the magnetization $M_N$ and the 
susceptibility $\chi$ on the noise parameter, obtained from simulations
on {\it directed} ER random graphs with $N=16000$ sites and several values of
connectivity $z$. In Fig. \ref{fig1a} each curve for $M_N$, for a given
value of $N$ and $z$, suggests that there is a phase transition from
an ordered state to a disordered state. The phase transition occurs at
a value of the critical noise parameter $q_{c}$, which is an
increasing function the connectivity $z$  of the {\it directed} ER
random graphs. In Fig. \ref{fig1b} we show the corresponding
behavior of the susceptibility $\chi$. The value of $q$ where $\chi$
has a maximum is here identified as $q_{c}(N)$ (Eq. \ref{eq9}). 

\begin{figure}[!htb]
\begin{center}
\includegraphics[angle=-90,scale=0.33]{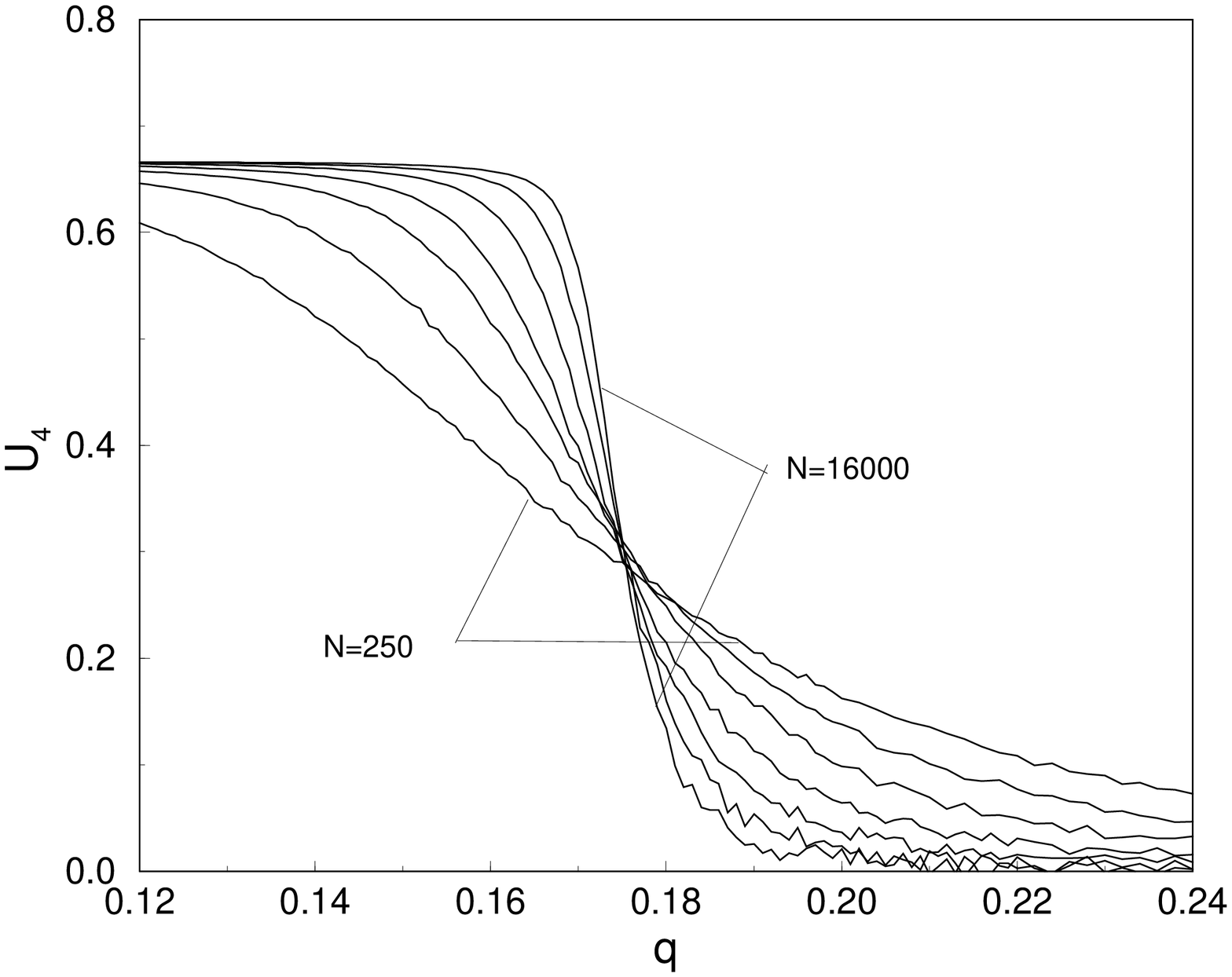}
\includegraphics[angle=-90,scale=0.33]{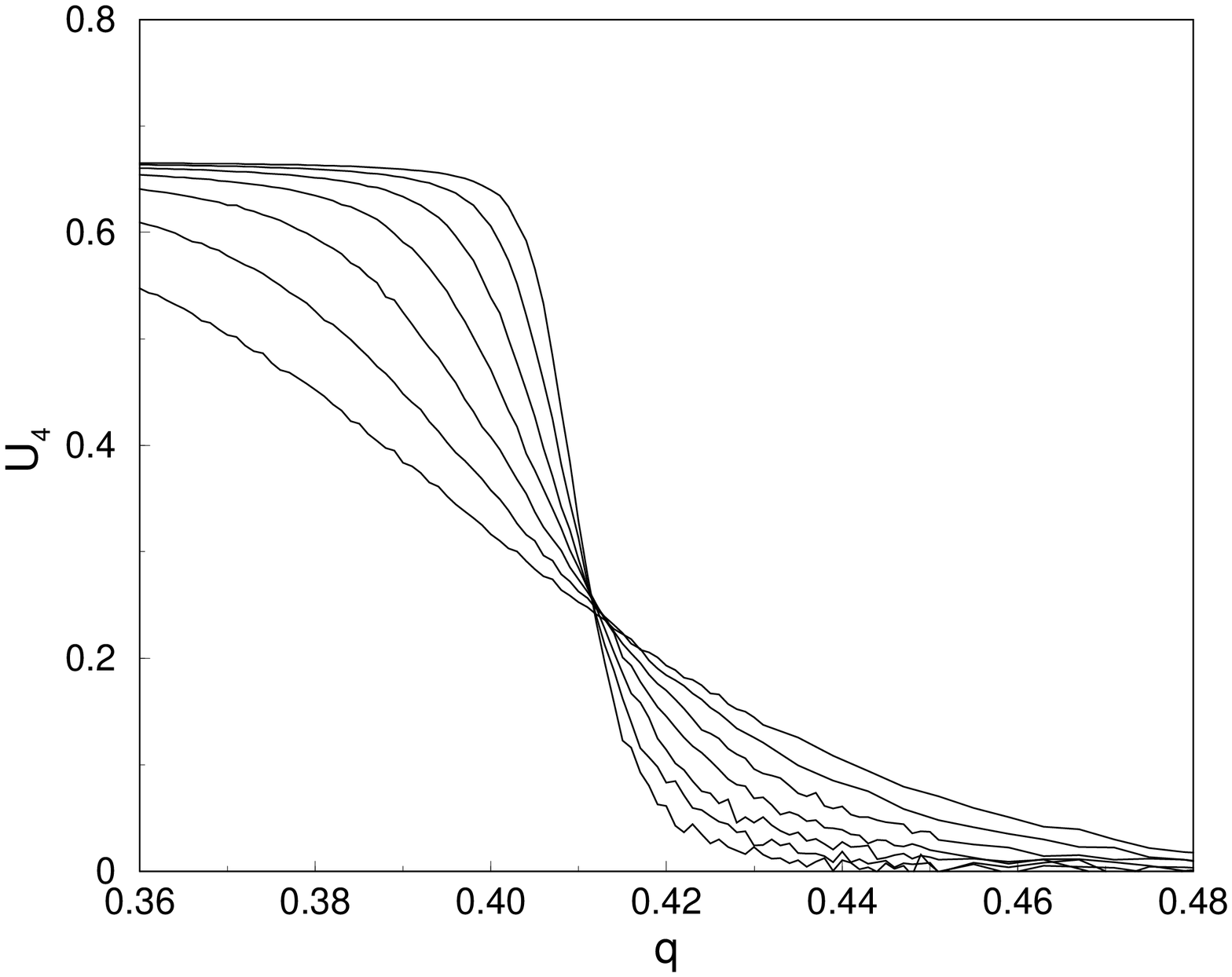}
\end{center}
\caption{Binder's fourth-order cumulant as a function of $q$ for several values of network sizes $N$ 
($N=250$, $500$, $1000$, $2000$, $4000$, $8000$ and $16000$), and for two different values of the 
connectivity $z$: $z=4$ (on the left-hand side) and $z=50$ (on the right-hand side).}
\label{fig2}
\end{figure}

To determine estimates for the critical point $q_{c}$, we calculate the Binder fourth-order 
magnetization cumulant $U$ at different values of the noise $q$ and several network sizes 
$N$. Finite size scaling predicts that for sufficiently large systems, these curves 
should have a unique intersection point $U^*$ \cite{binder}. The value of $q$ where 
this crossing occurs is the value of the critical noise $q_c$ which is not biased by 
any assumptions about critical exponents, since by construction, the Binder cumulant 
presents zero anomalous dimension, therefore it respects the correct critical behavior 
of the system near $q_c$ \cite{binder,pmco}.
\begin{figure}
  \centering
  \subfigure[]{\label{fig3a}%
    \includegraphics[angle=-90,width=0.48\textwidth]{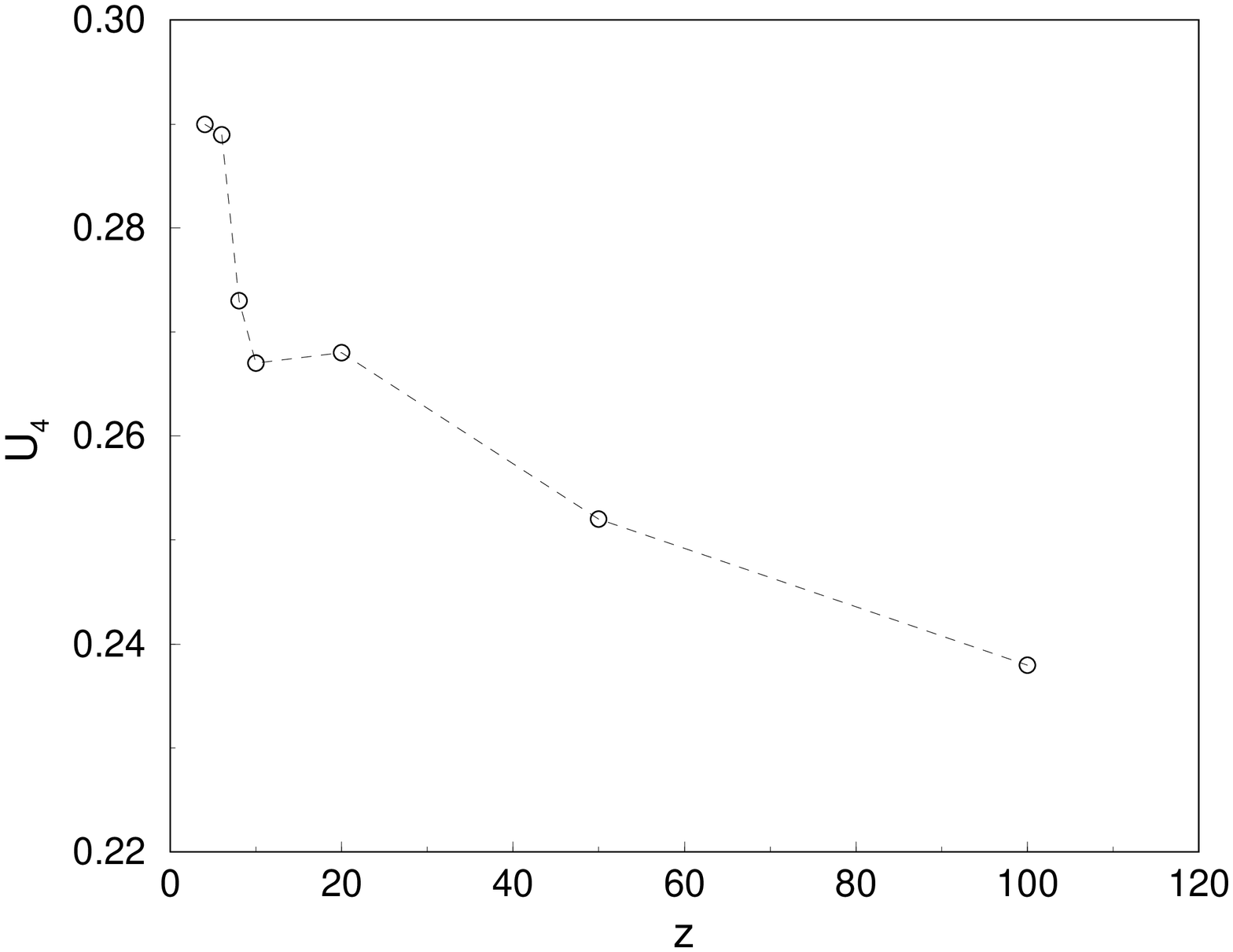}}%
  \quad%
  \subfigure[]{\label{fig3b}%
    \includegraphics[angle=-90,width=0.48\textwidth]{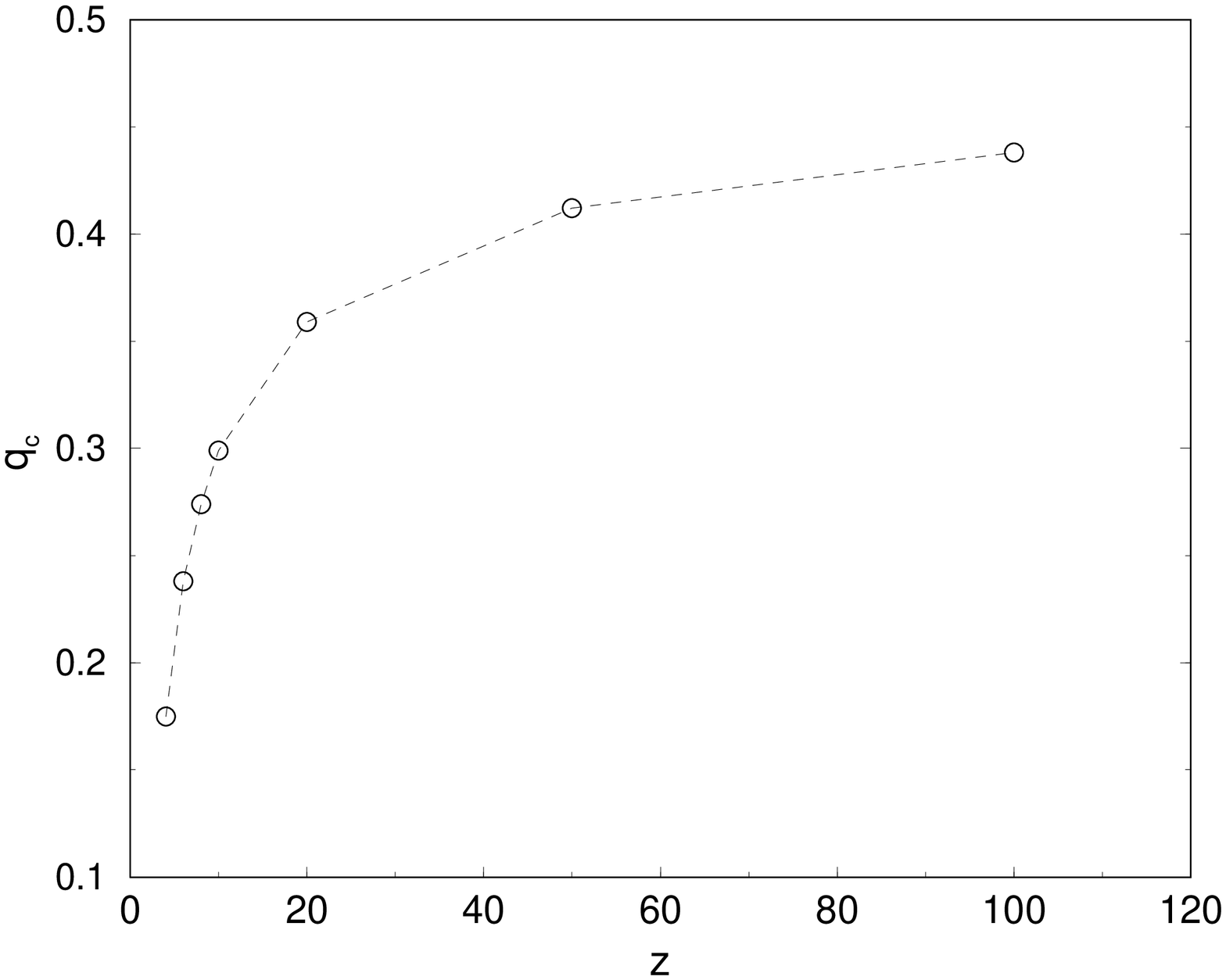}}
  \caption{{\bf (a)} The intersection point $U^*$ versus the connectivity $z$; {\bf (b)} Phase 
diagram: the critical values of the noise parameter $q_c$ as a function of the connectivity $z$.}
  \label{fig3}
\end{figure}

In Fig. \ref{fig2}  we plot the Binder's fourth-order cumulant for different values of $N$ and two
different values of $z$ ($z=4$ and $z=50$). The critical noise parameter $q_{c}$, for a
given value of $z$, is estimated as the point where the curves for different system sizes 
$N$ intercept each other. As we can noticed, there exists some dependence of the critical 
noise $q_c$ with the connectivity $z$, i.e., when the connectivity $z$ increases, the 
critical noise also increases: for $z=4$, $q_{c} \approx 0.18$ (left-hand side of 
accept Fig. \ref{fig2}), and for $z=50$, $q_{c} \approx 0.412$ (right-hand side of Fig. \ref{fig2}).
This same procedure is also made for several values of connectivity $z$, in this way the Binder's 
cumulant crossing point $U^\star$ as a function of the connectivity $z$ could be obtained 
(Fig. \ref{fig3a}), as well as the phase diagram is built, i.e., the dependence of the critical 
noise $q_c$ with the node connectivity $z$, what is shown in Figure \ref{fig3b}. Based on the 
fact that the crossing point of the cumulant (for different system sizes) gives the
transition point and the value of the cumulant $U^\star$ at the transition point indicates the 
universality class of the transition \cite{binder}. So, from Fig. \ref{fig3a}, which shows 
that the intersection 
point $U^\star$ of the cumulant decreases when the connectivity $z$ increases, it can be argued 
the present model does not fall into the same universality class of the equilibrium Ising model, 
since $U^{\star}=2/3$ well inside the ordered phase and $U \rightarrow 0$ well within the disordered 
phase. The Fig \ref{fig3b} indicates the increase of the critical noise $q_c$ with the increase 
of the node degree $z$, and moreover, for higher values of $z$, the critical noise $q_c$ approaches 
to that one obtained ($q_{c} \approx 0.5$) from mean-field theory when $z \rightarrow \infty$ 
\cite{pereira}. In addition, according to the mean-field theory \cite{pereira}, in the 
limit the connectivity $z$ goes to infinity, the magnetization approaches also to the 
value $0.5$, what one could infer from Fig \ref{fig1}. It is important to emphasize 
that these results obtained from our Monte Carlo simulations are in a good agreement with 
a previous work of the majority-vote model with noise on random graphs \cite{pereira}, 
in which simulations and mean-field analysis were performed to characterize the system 
transition.

\begin{figure}[!hbt]
 \centering
  \subfigure[]{\label{fig4a}%
    \includegraphics[angle=-90,width=0.48\textwidth]{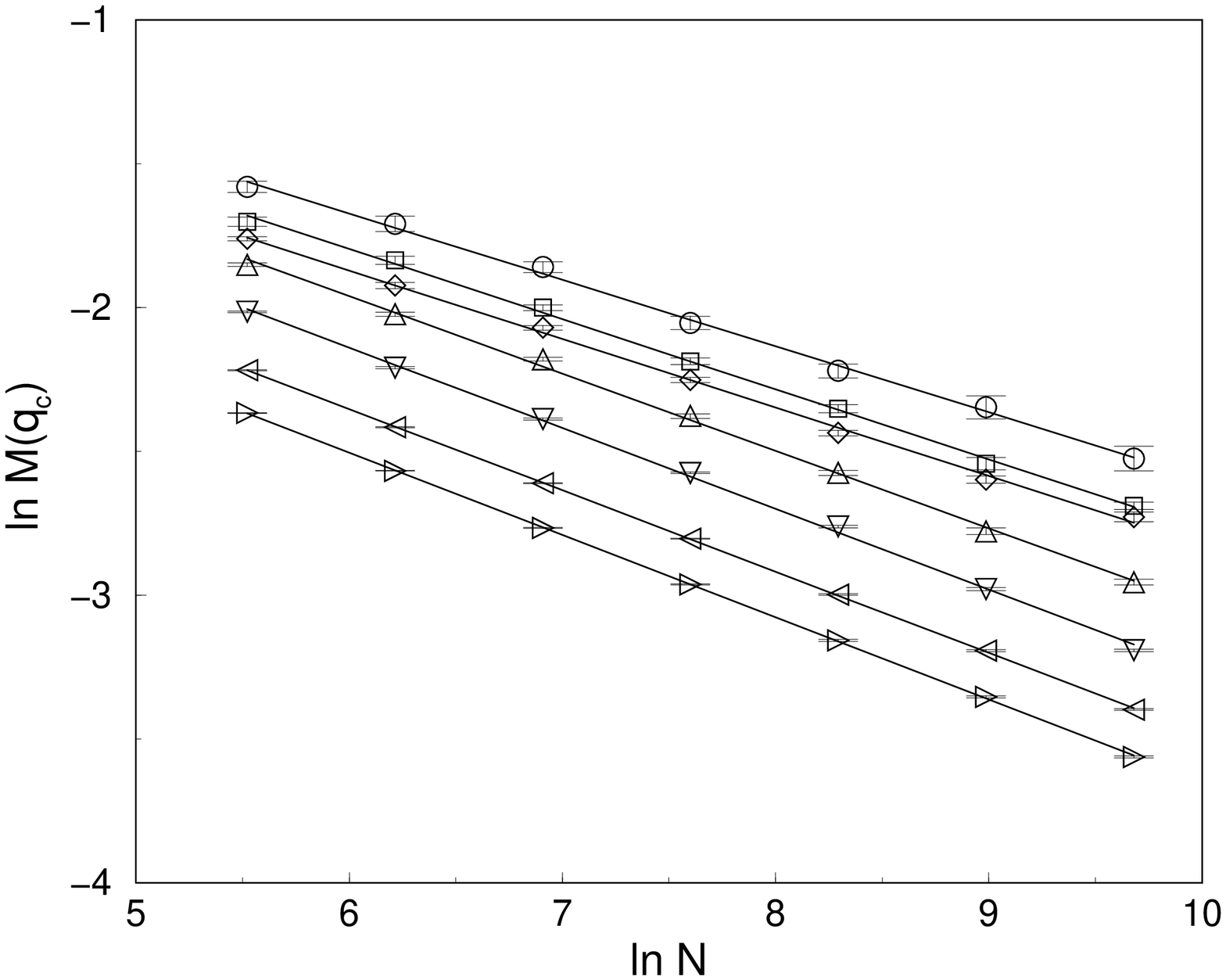}}%
  \quad%
  \subfigure[]{\label{fig4b}%
    \includegraphics[angle=-90,width=0.48\textwidth]{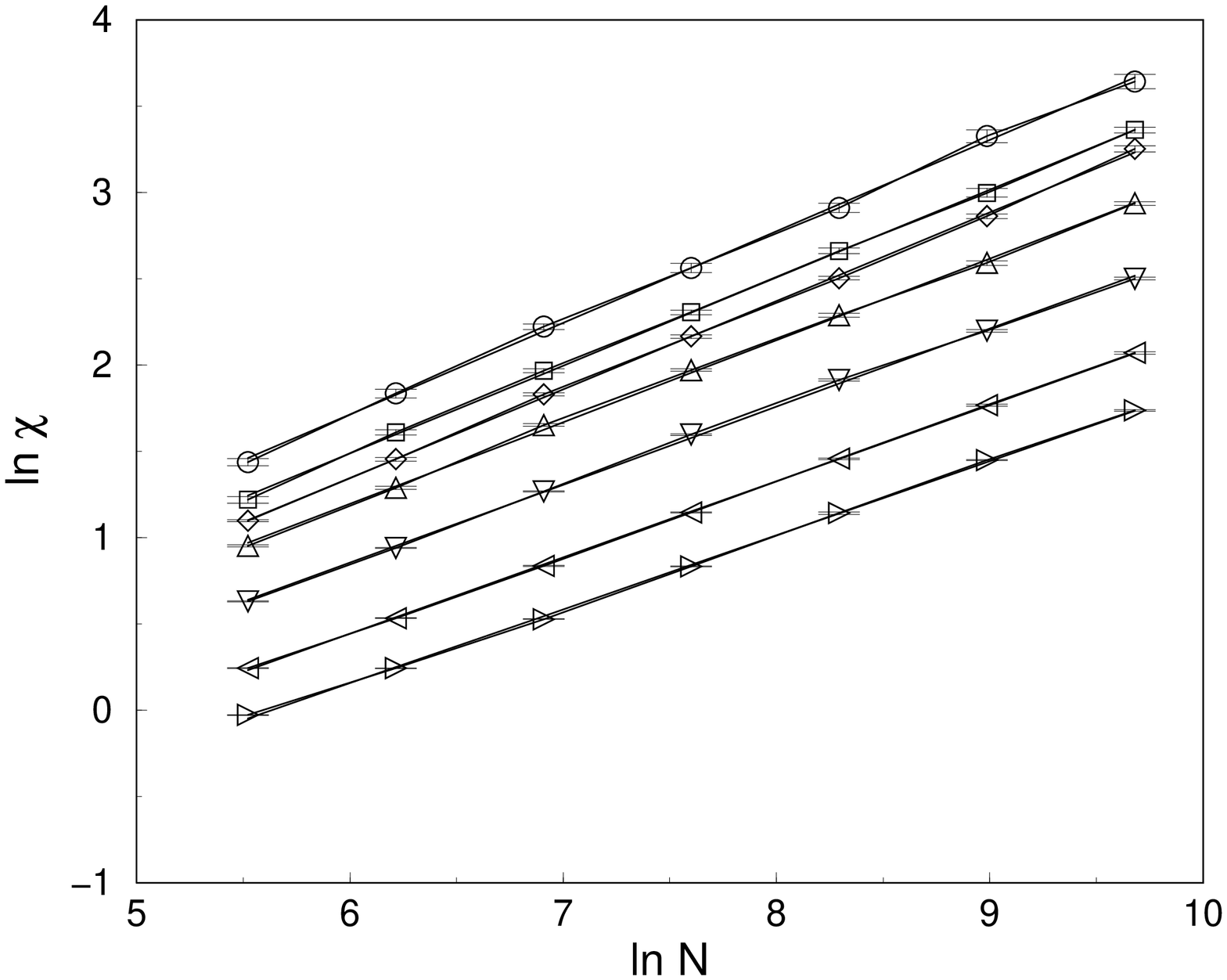}}
\caption{{\bf (a)} ln $M(q_{c})$ versus ln $N$ and {\bf (b)} ln $\chi(q_{c})$ 
versus ln $N$. In both figures, from top to bottom, $z=4$, $6$, $8$, $10$, $20$, $50$ and $100$. }
\end{figure}

Figure \ref{fig4a} and Figure \ref{fig4b} show, at $q=q_{c}$, the dependence of the
magnetization and the susceptibility with the system size $N$,
respectively, when different values of the connectivity $z$ are
considered. As we can see, the obtained straight lines, whose slopes
correspond to the exponents ratio $\beta/\nu$ (Fig. \ref{fig4a}) and $\gamma/\nu$
(Fig. \ref{fig4b}), confirm the scaling equations, given by Eqs. \ref{eq5} and
\ref{eq6}. Moreover, while there is a slight tendency for the exponent
$\beta/\nu$ to increase with $z$, the opposite occurs to the exponent
$\gamma/\nu$ that decreases with $z$.

\begin{figure}[!hbt]
\begin{center}
\includegraphics[angle=-90,scale=0.60]{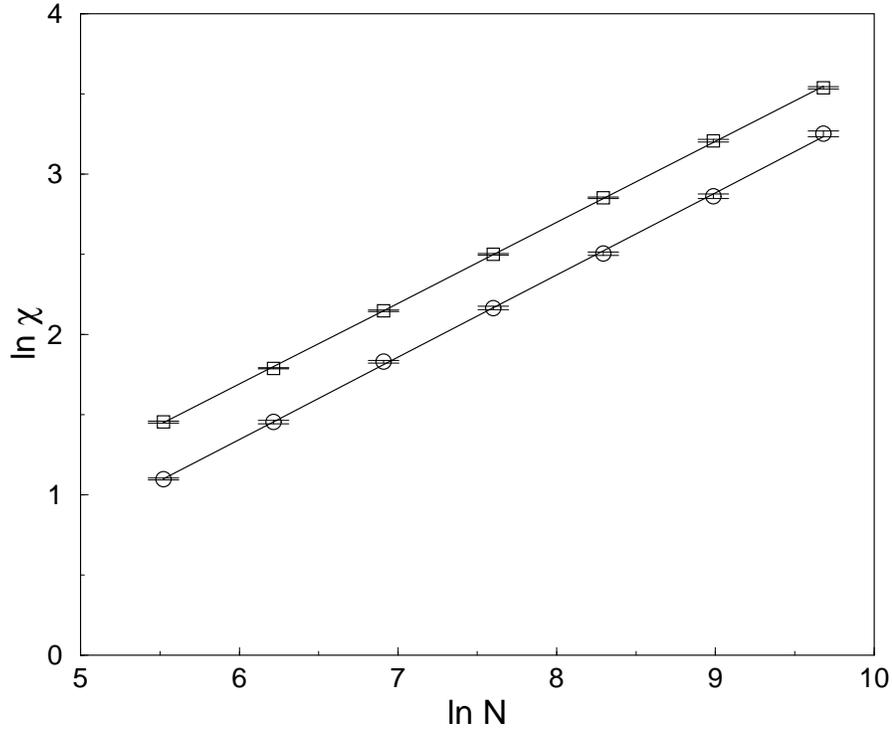}
\end{center}
\caption{Log-log plot of the susceptibility at its maximum $\chi_{\rm max}$ 
(squares) and $q = q_c$ (circles) versus $N$, for connectivity $z = 8$.}
\label{fig5}
\end{figure}

In Fig. \ref{fig5}, we present the dependence, at $q = q_c$, of the
susceptibility $\chi$ and of its maximum value $\chi_{\rm max}$ 
with the system size $N$. The exponent ratio $\gamma/\nu$ is obtained 
from the slopes of the straight lines. For almost all the values of $z$, 
the exponents $\gamma/\nu$ of the two estimates agree within error 
bars (see Table \ref{table1}). As already observed in Fig. 
\ref{fig4b}, an increased $z$ means a slight tendency to decrease 
the exponent ratio $\gamma/\nu$.

\begin{table}[!htb]
\begin{center}
\begin{tabular}{|c c c c c c c|}
\hline\hline
$ z $ & $q_{c}$ & $\beta/\nu$& ${\gamma/\nu}^{q_{c}}$ & ${\gamma/\nu}^{q_{c}(N)}$ & $ 1/\nu$ & $ D_{\rm eff} $ \\
\hline
$ 4 $ & $ 0.175(4) $ & $ 0.230(5) $ & $ 0.530(6) $ & $ 0.516(2) $ & $ 0.545(26)$ & $ 0.990(7)$\\
$ 6 $ & $ 0.238(3) $ & $ 0.243(4) $ & $ 0.509(4) $ & $ 0.511(2) $ & $ 0.488(16) $ & $ 0.995(5)$\\ 
$ 8 $ & $ 0.274(3) $ & $ 0.238(4) $ & $ 0.512(4) $ & $ 0.504(2) $ & $ 0.548(14) $ & $ 0.988(5)$\\
$ 10$ & $ 0.299(2) $ & $ 0.268(4) $ & $ 0.473(5) $ & $ 0.495(1) $ & $ 0.487(10) $ & $ 1.009(6)$\\
$ 20 $ & $ 0.359(2) $ & $ 0.280(4) $ & $ 0.451(4) $ & $ 0.485(2) $ & $ 0.510(10) $ & $ 1.011(5)$\\
$ 50 $ & $ 0.412(2) $ & $ 0.282(3) $ & $ 0.441(2) $ & $ 0.466(5) $ & $ 0.484(11) $ & $ 1.005(3)$\\
$ 100 $ & $ 0.438(2) $ & $ 0.286(2) $ & $ 0.428(3) $ & $ 0.440(8) $ & $ 0.520(19) $ & $ 1.000(3)$\\
\hline\hline
\end{tabular}
\end{center}
\caption{The critical noise $q_{c}$, and the critical exponents, 
for {\it directed} ER random graphs with connectivity $z$. Error bars
are statistical only.} 
\label{table1}
\end{table}

The Table \ref{table1} summarizes the values of $q_{c}$, the three critical
exponents $\beta/\nu$, $\gamma/\nu$, $1/\nu$ and the effective
dimensionality of the system. For all values of $z$, $D_{\rm eff}=1$,
which has been obtained from the hyperscaling hypothesis (Eq.10), 
therefore when $\beta/\nu$ increases, $\gamma/\nu$ decreases at
$q_{c}$, thus providing $D_{\rm eff}=1$ (along with errors). 

It is worth to mention that while our results for the majority-vote
model on a {\it directed} ER random graph are in good agreement with
the model on other complex networks, they are different from those
found for the majority-vote model and the traditional Ising model on a 
square lattice \cite{mario}. Moreover, the critical exponents and the
hyperscaling relation obtained from our simulations corroborate a
proposition that the majority-vote models defined on a regular
lattice, on small-world networks \cite{campos}, on Voronoy-Delaunay
lattice \cite{lima0}, on Barab\'asi-Albert networks
\cite{lima1,lima2}, and on {\it undirected} Erd{\H o}s--R\'enyi's random
graphs \cite{pereira}, belong to different universality
classes.

\section{Conclusion}

Using Monte Carlo simulations, we found a second-order phase
transition of the majority-vote model with noise on {\it directed} ER 
random graphs. The observed phase transition, which occurs with
connectivity $z>1$, was completely characterized through the phase
diagram and the critical exponents. Although the obtained exponents
are different from those for the same model on other topologies 
\cite{mario,campos,lima0,lima1,lima2,lima3}, they suggest that
the majority-vote models defined on a regular lattice, on small-world 
networks, and on Erd{\H o}s--R\'enyi's random graphs, belong to
different universality classes. In fact, it was found that the 
critical exponents depend on the long-range interactions (short cuts), 
i.e., they depend on the node degree $z$, in such way that results in a 
larger robustness of the system against the noise: the larger the connectivity 
$z$, the greater the critical noise $q_c$. The found values $U^\star$ of 
the Binder's fourth-order cumulant at the transition point for different 
values of the connectivity $z$ were different from those obtained for the 
isotropic majority-vote model on a regular square lattice and, moreover, for the 
square Ising model, what corroborates the conjecture that the majority-vote 
model on complex network does not belong to the same universality class of 
the equilibrium Ising model. Finally, we should remark that the importance of 
our results is not only in presenting a set of numbers, but it is in characterizing 
completely the found phase transition, i.e., to determine the whole set of 
the critical exponents of the order-disorder phase transition appearing in 
the system. Moreover, the full set of critical exponents for several values 
of the connectivity $z$, which is really different from the exact values of 
the Ising model and majority-vote model on a regular lattice, shows clearly 
the influence of the small-word effects in the majority-vote model, as well 
as it points out the dependence of the critical exponents with the connectivity 
$z$. Therefore, the well-known effects of the shortest path length in systems 
with small-world topology on the information propagation are the responsible for 
making more effective the influence of the neighborhood, which now is not 
constrained by the geographical distance between the individuals, as observed in 
regular lattices. Of course, we are aware that random graphs are seldom very 
good models for the type of networks one finds in nature \cite{ba,moreno}. 
While they have a low shortest path length, they do not have another important 
property of most observed networks: clustering. This is obviously not the case 
in our studied topology, where all vertex pairs are independently connected with 
the same probability.

\section{Acknowledgments}
%The authors thanks  D. Stauffer for many suggestions and fruitful 
%discussions during the development this work and also for the revision of
%this paper. 
F.W.S. Lima acknowledges the Brazilian agency FAPEPI
(Teresina-Piau\'{\i}-Brasil) for  its financial support. A.O. Sousa 
is grateful to the Brazilian Foundation FAPERN for finantial support. 
This work also was supported the system SGI Altix 1350 the computational 
park CENAPAD.UNICAMP-USP, SP-BRAZIL.


\begin{thebibliography}{00}

\bibitem{gray} L. Gray, in {\em Particle Systems, random Media and Large 
Deviations}, ed. R. Durrett, American Mathematical Society (Providence 
-- Rhode Island) 1985, p. 149.

\bibitem{mario} M.J. Oliveira, J. Stat. Phys. {\bf 66}, 273 (1992).

\bibitem{santos} M. A. Santos, S. Teixeira, J. Stat. Phys. {\bf 78}, 
963 (1995).

\bibitem{crochik} L. Crochik, T. Tom\'e, Phys. Rev. E {\bf 72}, 
057103 (2005).

\bibitem{ising} W. Lenz, Z. Phys. {\bf 21}, 613 (1920); 
E. Ising, Z. Phys. {\bf 31}, 253 (1925); M. Hasenbusch, Int. J. 
Mod. Phys. C {\bf 12}, 911 (2001).

\bibitem{critical} J. J. Binney, N. J. Dowrick, A. J. Fisher, and 
M. E. J. Newman, {\em A theory of critical phenomena. An Introduction 
to the renormalization group}, Clarendon Press (Oxford) 1992.

\bibitem{grin} G. Grinstein, C. Jayapralash, and Yu He, Phys. Rev. 
Lett. {\bf 55}, 2527 (1985).

\bibitem{none} M.J. de Oliveira, J.F.F. Mendes, and M.A. Santos, J. Phys. 
A {\bf 26}, 2317 (1993); P. Tamayo, F.J. Alexander, and R. Gupta, Phys. 
Rev. E {\bf 50}, 3474 (1995); T. Tom\'e´ and J.R. Drugowich de Fel\'icio, 
Phys. Rev. E {\bf 53}, 3976 (1996); N.R.S. Ortega, C.F. Pinheiro, T. Tom\'e 
and J.R. Drugowich de Fel\'icio, Physica A {\bf 255}, 189 (1998).

\bibitem{campos} P.R. Campos, V.M. Oliveira, and F.G.B.Moreira, Phys. 
Rev. E {\bf 67},026104 (2003).

\bibitem{lima0} F.W.S. Lima, U.L. Fulco, and R.N. Costa Filho, Phys. 
Rev. E {\bf 71}, 036105 (2005).

\bibitem{lima1} F.W.S. Lima, for Int. J. Mod. Phys. C {\bf 17}, 
1257.

\bibitem{pereira} L.F.C. Pereira and F.G. Brady Moreira, Phys. Rev. E {\bf 71},
016123 (2005).

\bibitem{watts}  D.J. Watts, S.H. Strogatz, Nature {\bf 393}, 440 (1998); M.E.J. 
Newman and D.J. Watts, Phys. Rev E {\bf 60}, 7332 (1999); M.E.J. Newman and D.J. 
Watts, Phys. Lett. A {\bf 263}, 341 (1999).

\bibitem{er} P. Erd{\H o}s and A. R\'enyi, Publ. Math. Debrecen {\bf 6}, 
290 (1959); P. Erd{\H o}s and A. R\'enyi, Publ. Math. Inst. Hung. Acad. 
Sci. {\bf 5}, 17 (1960); P. Erd{\H o}s and A. R\'enyi, Bull. Inst. Int. 
Stat. {\bf 38}, 343 (1961).

\bibitem{boll} B. Bollob\'as, Random Graphs, Academic Press (New York) 1985.

\bibitem{VD} F. W. S. Lima, J. E. Moreira, J. S. Andrade, Jr., 
and U. M. S. Costa, Physica A {\bf 283}, 100 (2000); F. W. S. Lima, 
U. M. S. Costa, M. P. Almeida, and J. S. Andrade, Jr., Eur. Phys. J. 
B {\bf 17}, 111 (2000).

\bibitem{ba} R. Albert and A.L. Barab\'asi, Rev. Mod. Phys. {\bf 74}, 
47 (2002)

\bibitem{alex} A. Aleksiejuk, J.A. Ho\l yst and D. Stauffer, Physica A 
{\bf 310}, 269 (2002).

\bibitem{jff} J.J.F. Mendes and M. A. Santos, Phys. Rev. E {\bf 57}, 
108 (1998)

\bibitem{social1} D. Staufer, J. Artificial Soc. Social Simulations 
{\bf 5}, 1 (2002); D. Staufer, Conf. 50th Anniversary of the Metropolis 
Algorithm (2003), in press, cond-mat/0307133; K. Sznajd-Weron and J. 
Sznajd, Int. J. Mod. Phys. C {\bf 11}, 1157 (2000); M.C. Gonzalez, 
A.O. Sousa and H.J. Herrmann, Int. Journal of Modern Physics C {\bf 15}, 
45 (2004). D. Stauffer, A.O. Sousa, C. Schulze, Journal of Artificial 
Societies and Social Simulation (2004), A.O. Sousa, Physica A {\bf 348}, 
701, (2005).Tu Yu-Song, A.O. Sousa, Kong Ling-Jiang and Liu Mu-Ren, 
Int. J. Mod. Phys. C {\bf 17}, (2005).

\bibitem{social2} G. Defuant, D. Neau, F. Amblard and G. Weisbuch, 
Adv. Complex Syst. {\bf 3}, 87 (2000); G. Weisbuch, G. Defuant, 
F. Amblard and J.-P. Nadal, Complexity {\bf 7}, 55 (2002); G. Defuant, 
F. Amblard, G. Weisbuch and T. Faure, J. Artficial Soc. Social Simulation
{\bf 5}, 1 (2002); R. Hegselmann and M. Krause, J. Artificial Soc. Social Simu-
lation {\bf 5}, 2 (2002); U. Krause, {\it Modellierung und Simulation von Dynamiken
mit vielen interagierenden Akteuren}, eds. U. Krause and M. St\"ockler 
(Bremen University,1997), p. 37.

\bibitem{galam1} S. Galam, J. Math. Psychology {\bf 30}, 426 (1986); 
S. Galam, J. Stat. Phys. {\bf 61}, 943 (1990); S. Galam, Physica A 
{\bf 285}, 66 (2000); S. Galam, Eur. Phys. J. B {\bf 25}, 403 (2002).

\bibitem{galam2} S. Galam, Europhys. Lett. {\bf 70}, 705 (2005). 

\bibitem{lama} M.S. de la Lama, J.M. Lopez, H.S. Wio, Europhys. Lett. {\bf 72}, 
851 (2005); H.S. Wio, M.S. de la Lama and J.M. Lopez, Physica A {\bf 371}, 108 (2006).

\bibitem{graph} Bollob\'as, {\it Random Graphs} Academic Press (London) 1985; 
B. Bollob\'as, {\it Modern Grapgh Theory}, Graduate Texts in Mathematicas, 
Springer (New York) 1998.

\bibitem{cluster} S. Wasserman, K. Faust,B. Bollob\'as, Social Networks Analysis, 
Cambridge University Press (Cambridge) 1994.

\bibitem{binder} K. Binder, Z. Phys. B 43, 119 1981; K. Binder, in {\it Finite Size 
Scaling and Numerical Simulation of Statistical Systems}, edited by V. Privman, 
World Scientific (Singapore) 1990, 174; D.P Landau and K. Binder, {\it A Guide to 
Monte Carlo Simulations in Statistical Physics}, Cambridge University Press (Cambridge) 
2000.

\bibitem{pmco} J.S. S\'a Martins and P.M.C. de Oliveira, Braz. J. Phys {\bf 34}, 1077 (2004). 

\bibitem{lima2} F.W.S. Lima, Commun. Comput. Phys., {\bf 2}, 358 (2007).

\bibitem{lima3} Edina M. S. Luz and F.W.S. Lima , for Int. J. Mod. Phys. C.

\bibitem{moreno} S. Boccaletti, V. Latora, Y. Moreno, M. Chavez, D.-U Hwang, Phys. Rep. 
{\bf 424}, 175 (2006).

\end{thebibliography}
\end{document}